\documentclass[sigconf,nonacm]{acmart}
\usepackage{booktabs}
\usepackage{graphicx}
\usepackage{adjustbox}
\usepackage{xurl}

\setcopyright{none}
\AtBeginDocument{\hypersetup{pdfcreationdate={D:19700101000001Z},pdfmoddate={D:19700101000001Z},pdfmetadate={D:19700101000001Z},pdfdate={1970-01-01}}}

\newcommand{\npanelNModels}{13}
\newcommand{\npanelNHostedOr}{9}
\newcommand{\npanelJevThirteenChoice}{0.210}
\newcommand{\npanelJevFourteenChoice}{0.037}
\newcommand{\nnWorldsPerFamily}{100}
\newcommand{\pfNItems}{800}
\newcommand{\pfNBoot}{2000}
\newcommand{\pmfNChange}{4}
\newcommand{\pmfNSystems}{13}
\newcommand{\prsNDev}{395}
\newcommand{\prsNTest}{405}
\newcommand{\prsRaw}{0.201}
\newcommand{\prsTemp}{0.163}
\newcommand{\prsNHurts}{5}
\newcommand{\prsLunaRaw}{0.260}
\newcommand{\prsNBagHelpsCI}{10}
\newcommand{\prsNTempLowersCI}{13}
\newcommand{\prsNSystems}{13}
\newcommand{\prsNHurtBoolWorse}{4}
\newcommand{\prsLunaBool}{0.306}
\newcommand{\pdeRegretConst}{0.016}
\newcommand{\pdeRegretLo}{0.024}
\newcommand{\pdeRegretHi}{0.071}
\newcommand{\pdeNEvBag}{4}
\newcommand{\pdeNEvStars}{4}
\newcommand{\pdeNEvChat}{5}
\newcommand{\pdeNBothWorseCI}{3}
\newcommand{\pdeNSystems}{13}
\newcommand{\pdeLunaB}{0.1}
\newcommand{\pdeLunaC}{16.5}
\newcommand{\pdeLunaK}{13.9}
\newcommand{\ppbLunaDevB}{0.00}
\newcommand{\ppbLunaDevC}{1.76}
\newcommand{\ppbLunaDevBSecond}{0.00}
\newcommand{\ppbLunaDevCSecond}{1.79}
\newcommand{\pbnOrc}{0.119}
\newcommand{\pbnOrcMinusB}{[\ensuremath{-}0.0022, \ensuremath{-}0.0015]}
\newcommand{\pbnNBetterChoice}{1}
\newcommand{\pbnSel}{0.121}
\newcommand{\pbnB}{0.120}
\newcommand{\pbnSelMinusB}{[0.0003, 0.0017]}
\newcommand{\pinNoInfoNames}{\texttt{open-lumma}}
\newcommand{\pinHostedRhoMin}{0.72}
\newcommand{\pinConstWrong}{13.7--13.9}
\newcommand{\pinNotBeatNames}{\texttt{open-lumma}}
\newcommand{\paAlpha}{3, 1.5, 0.75, 0.375}
\newcommand{\paNBags}{100}
\newcommand{\paNTest}{50}
\newcommand{\paRWrong}{12.8}
\newcommand{\paNChoiceWorse}{3}
\newcommand{\pdUltraGapChangeRep}{\ensuremath{-}0.022 [\ensuremath{-}0.033, \ensuremath{-}0.012]}
\newcommand{\pdDays}{11}
\newcommand{\pdDecUltraRepeat}{4.2}
\newcommand{\pdNSystems}{9}
\newcommand{\pnNSystems}{9}
\newcommand{\pnNItems}{800}
\newcommand{\pnSelMae}{0.184}
\newcommand{\pnBoolMae}{0.184}
\newcommand{\pnOrcMae}{0.176}
\newcommand{\pnOrcMinusBool}{[\ensuremath{-}0.0084, \ensuremath{-}0.0069]}
\newcommand{\pnSelMinusBool}{[\ensuremath{-}0.0001, 0.0016]}

\newcommand{\sys}[1]{\texttt{#1}}

\title{Typed Decision Probabilities Can Fail Silently}
\author{Junwei Yao}
\email{yao.junw@northeastern.edu}
\affiliation{\institution{Independent Researcher}\country{USA}}
\author{Yongxi Zhou}
\affiliation{\institution{Independent Researcher}\country{USA}}
\author{Wenbo Ye}
\affiliation{\institution{Independent Researcher}\country{USA}}
\author{Yanxian Chen}
\affiliation{\institution{Independent Researcher}\country{USA}}
\author{Jianan Liu}
\affiliation{\institution{Independent Researcher}\country{USA}}
\author{Zike Yang}
\affiliation{\institution{Independent Researcher}\country{USA}}

\begin{document}
\begin{abstract}
Typed decision APIs let a program ask a language model a yes/no question (a \emph{Boolean}) or a multiple-choice
question (a \emph{Choice}) and get back a probability for each answer, which developers threshold to make decisions.
A developer who does this may rely on three assumptions: acting on these probabilities beats ignoring the input and
using the base rate; once measured, the more accurate question type stays the more accurate one on a new task, model
version and run; and a check or fix that works on test questions with known answers also works on the developer's own
task. We ask whether reliability measured in one setting carries over to the next, testing each assumption on $\npanelNModels$ released systems, on synthetic tasks where the correct probability is known and
on human judgments from ChaosNLI. Each assumption fails on some systems. On the main synthetic tasks, for $\pdeNBothWorseCI$ of the
$\npanelNModels$ systems, one of them a hosted service, decisions from either question type are measurably wrong more
often than decisions from the base rate. For some systems the more accurate type changes from one task to another, and
for one hosted service it reverses between runs under the same model name. Answers to per-option questions that sum far from one must be inaccurate, but choosing between per-option yes/no and
two-way questions by this sum does not measurably improve on always using yes/no. A fix that rebuilds the Choice from the Booleans helps on
synthetic bags of colored tokens but makes some systems worse on human judgments, where fitting to labeled data
helps. Every probability returned is well-formed, so none of these failures shows in a single answer. Developers should
measure the question type they use on their own task, and measure it again when the model version or the run changes.
\end{abstract}



\maketitle

\section{Introduction}

\emph{Typed decision interfaces} let a program ask a language model a typed question about an input text (the
\emph{state}) and receive a probability for each answer. A \emph{Boolean} asks yes or no; a \emph{Choice} picks one of
several named options. TypeSafe AI introduced the interface with its Jev models~\citep{svc2026docs}, and OpenAI's
Decisions API~\citep{openai2026decisions}, decision models served through OpenRouter, and open
models~\citep{hf2026decider,hf2026bosun,hf2026lumma} followed. The interface normalizes each question on its own:
nothing ties the Boolean ``Is it red?'' to the probability a Choice over colors puts on red. A caller may have measured
only one type, or may plan to rebuild a Choice from Booleans. It then needs to know whether the other type is as
accurate, and whether the rebuild helps.

Concurrent audits compare the types. With label-free checks on ChaosNLI and PubMedQA items, \citet{arxiv2609_33209} find violations of the probability
axioms, including disagreement between a Boolean and the matching Choice. \citet{arxiv2610_11392} report that re-posing one Boolean question
as a two-option Choice gives nearly the same probabilities on every model they test. On four-cell
worlds with exact posteriors, \citet{arxiv2609_37470} find that the favored interface differs between an open typed
model and a commercial one; they also evaluate decisions and the average of the two types. \citet{arxiv2609_35342}
poses 92 problems with known probability through the Boolean, the Choice and the ordered Score type of one model. It
finds the Choice least well calibrated and repairs it with a learned remapping. Disagreement shows that at least one answer is wrong,
not which.

Consider a developer who moderates comments by asking such an interface whether a comment is abusive, and deletes the
comment when the probability exceeds $0.5$. The developer may rely on three assumptions that the interface does not guarantee. (A1) Acting on the probabilities beats ignoring the comment and acting on how common abuse is overall, the
\emph{base rate}. (A2) Once the developer has measured which question type is more accurate, it stays the more
accurate one on the next task, after the next model update and on the next run. (A3) A check or fix that needs no
labeled examples, and works on synthetic test questions with known answers, also works on the developer's own task.

We test each assumption on $\npanelNModels$ released systems, on synthetic tasks where the correct probability is
known and on human judgments from ChaosNLI. Each assumption fails on some systems. For A1, decisions from either
question type are detectably wrong more often than decisions from the base rate on $\pdeNBothWorseCI$ of
$\pdeNSystems$ systems. For A2, the more accurate type changes from one kind of task to another on $\pmfNChange$
systems, and for one hosted system it detectably reverses between runs under the same model name. For A3, choosing between per-option Booleans and per-option two-way Choices by whether their answers sum to one does
not detectably improve on always using the Boolean, on the
synthetic bags of colored tokens or on ChaosNLI. Rebuilding the Choice from the Booleans detectably lowers the error of $\prsNBagHelpsCI$ of
$\prsNSystems$ systems on the bags, but raises it for $\prsNHurts$ systems on human judgments. These failures are silent. Every probability returned is well-formed, and nothing in a single
answer shows that the type, the version or the run is the unreliable one. The three assumptions share one question: whether reliability measured in one setting supports the next choice of
question type or repair. The setting can be one task, model version or run, or synthetic test questions with known answers. Three of the findings do not follow from earlier results on interface disagreement and calibration. On the same
items, the question type decides whether threshold decisions beat the base rate, and for some systems neither type
does. Under an unchanged model name, the more accurate type reverses between runs. The same label-free repair that
helps on the synthetic bags makes some systems worse on human judgments.

\section{Setup}\label{sec:setup}

\paragraph{Systems} \sys{gpt-6-luna} is served through OpenAI's Decisions API. Another $\npanelNHostedOr$
hosted models are served through OpenRouter's typed-decision endpoint, with provider fallbacks disabled; two of them
are versions of one model (\sys{jev-1.13}, \sys{jev-1.14}). Three open typed models (\sys{open-decider}, \sys{open-bosun}, \sys{open-lumma}~\citep{hf2026decider,hf2026bosun,hf2026lumma};
nominally 2B, 1.7B and 0.6B parameters) run from their released weights and inference code. Every system is used as released (Appendix~\ref{app:protocol}). All $\npanelNModels$ systems enter every analysis
except four that use the $\pdNSystems$ OpenRouter systems: the re-measurement, the bags with unequal base rates, and
the partition check as a selector, on the bags and on ChaosNLI.

\paragraph{Tasks with exact targets} Each of three task families has $\nnWorldsPerFamily$ items whose event
probabilities follow from counts stated in the text, so the exact answer is known. A \emph{bag} reads, for example,
``An opaque bag holds 12 red, 5 green, 2 blue, 1 yellow tokens (20 in total). One token was drawn uniformly at random
and its color has not been revealed.'' The exact probability that the drawn token is red is $12/20=0.6$. Each bag is
asked four per-color Booleans (``Is the drawn token red?'') and a four-way Choice over the colors, among other
questions. Unless stated otherwise, \emph{Choice} means this original multi-option question. A \emph{histogram} states a product's 1--5 star distribution. A \emph{chat log}
states the turn at which customers first became angry. Each is asked a Choice over the levels and Booleans about
cumulative events (``at least $k$ stars'', ``angry by turn $k$''). A state's questions go in one request unless Appendix~\ref{app:protocol} states
otherwise. On one hosted typed model, questions packed into one call do not interact~\citep{arxiv2610_16655}.

\begin{table}[t]\centering\footnotesize\setlength{\tabcolsep}{4pt}
\begin{tabular}{@{}lrrrrr@{}}
\toprule
 & red & green & blue & yellow & MAE \\
\midrule
exact probability & 0.60 & 0.25 & 0.10 & 0.05 & -- \\
Booleans, $P(\text{yes})$ & 0.70 & 0.30 & 0.20 & 0.10 & 0.075 \\
four-way Choice & 0.90 & 0.05 & 0.03 & 0.02 & 0.150 \\
rebuilt Choice (Booleans $\div\,1.3$) & 0.54 & 0.23 & 0.15 & 0.08 & 0.040 \\
\bottomrule
\end{tabular}
\caption{The bag of Section~\ref{sec:setup} asked as four Booleans and as a four-way Choice, with illustrative answers
that come from no system. The error ratio is $0.150/0.075=2$, so the Boolean is the more accurate type. The Booleans
sum to $1.3$, a deviation of $0.3$, so their MAE is at least $0.3/4=0.075$; here it is exactly that. At $\tau=0.5$,
all three answers act on red, as the exact probability does.}\label{tab:example}
\end{table}

\paragraph{Per-type error} An answer's error is its distance from the exact probability $p^*$. If the Boolean ``Is
the drawn token red?'' returns $P(\text{yes})=0.7$ for the bag above, its error is $|0.7-0.6|=0.1$; a Choice is scored
the same way on each option. MAE averages these errors over events and items; Table~\ref{tab:example} works through one bag. Both types are scored on the same
events. On the bags these are the four colors. On histograms and chat logs they are the cumulative events, and the
Choice's probability for such an event is the sum over the options it covers. The \emph{error ratio} is the Choice's MAE divided by the Boolean's. Above one, the
Boolean is more accurate; we call the type with the lower MAE the \emph{more accurate type}. Intervals are 95\%
bootstrap intervals, per comparison and unadjusted. A difference is \emph{detectable} when its interval excludes zero
(one, for a ratio); counts of systems refer to detectable differences.

\paragraph{Decisions} A caller often acts on a probability by thresholding it: it acts when the probability exceeds
$\tau$. We call a decision \emph{wrong} when it differs from the decision the exact probability would give. That happens when
the answer and the exact probability fall on different sides of $\tau$; it says nothing about which token was drawn. With $\tau=0.5$, answering $0.4$ for the red token of the bag above is a wrong decision, because the exact
probability $0.6$ would act. We report the share of wrong decisions over the \emph{matched events}, the events that both types answer (each color
of a bag, each cumulative event of a histogram or chat log), and $\tau\in\{0.1,0.3,0.5,0.7,0.9\}$. Every decision counts once, so each item of a bag, a histogram and a chat log
contributes $\pdeNEvBag$, $\pdeNEvStars$ and $\pdeNEvChat$ events. A type can lose to the base rate in two ways: its answers ignore the input, or they
follow it on the wrong scale. To tell these apart, we also report each type's rank correlation with the exact probability, within one event and
across items. We also report its decisions after a temperature fitted on half of the bags.

\paragraph{Base-rate rule} The baseline ignores the input text and gives every Choice option the same probability (the uniform constant of Table~\ref{tab:inform}). On
the bags it therefore answers $1/4$ for every color. On histograms and chat logs, we sum these uniform option
probabilities for each cumulative event, as we do for a model's Choice. Its answers are thresholded like a model's. At
each of the five thresholds, this rule makes the same decisions as a constant at each event's mean exact probability
over items, which would need the labels. On a second set of bags with unequal colors, the base rate is instead
estimated on half of the bags (Appendix~\ref{app:protocol}).

\paragraph{Re-measurement} The $\pdNSystems$ OpenRouter systems were asked the same items again about $\pdDays$ days after
the first run, under the same model identifiers. An
interval from one run covers the choice of items but not the variation between runs, which needs at least two
runs~\citep[cf.][]{zhou2026repeated}.

\paragraph{Partition check} When each option of a partition is asked as its own question, the answers should sum to
one, and how far they deviate needs no labels. For the bag above, Booleans of 0.7, 0.3, 0.2 and 0.1 for the four colors
sum to 1.3, a deviation of 0.3. The deviation bounds the error from below: since $|\sum_l q_l-1|\le\sum_l|q_l-p_l|$, a
deviation $D$ over $k$ options implies an MAE of at least $D/k$, here $0.075$. A large deviation therefore certifies a
large error, but a small one certifies nothing, just as disagreement between two interfaces certifies only part of
their error~\citep{arxiv2609_37470}. Here the options are asked as Booleans or as \emph{per-option two-way Choices} (on a bag, ``The drawn token is $X$.'' against ``The drawn token is a colour other than $X$.''). The original Choice sums to one up to rounding, so the check applies only to these per-option forms. Used as a
selector, the check picks for each system whichever of the two deviates less on average, and the Boolean on a tie.

\paragraph{Repairs on human labels} ChaosNLI~\citep{nie2020chaosnli} gives 1,599 MNLI pairs a label distribution from
100 annotations each. On $\pfNItems$ of them, split into development and test halves, the $\prsNSystems$ systems answer
a three-way Choice and one Boolean per label; an answer's error is its distance from the label distribution. We compare
two repairs of the \emph{original Choice}, the Choice as returned (three-way on ChaosNLI). The \emph{rebuilt Choice} replaces it with the per-label Booleans divided by
their sum. For the bag above, each of 0.7, 0.3, 0.2 and 0.1 would be divided by 1.3. It has no fitted parameters.
\emph{Temperature scaling} divides the log-probabilities of the original Choice by a constant fitted on development labels;
its search range includes $T=1$, no scaling.

\section{Results}\label{sec:results}

\subsection{A1: Which type, if any, beats the base rate?}

A caller that thresholds a type's probabilities needs that type to decide better than a rule that ignores the input.
This section asks whether either type does.

\paragraph{The choice of type matters} Choosing the type can decide whether decisions beat a rule that ignores the
input text. On \sys{gpt-6-luna}, across the three task families, $\pdeLunaB$\% of the Boolean's threshold decisions are
wrong and $\pdeLunaC$\% of the Choice's; the base-rate rule is wrong on $\pdeLunaK$\%. For this system the Boolean
beats the baseline and the Choice loses to it. A caller who measured only the Choice would find the system worse than
ignoring the input; one who measured only the Boolean would find its decisions almost always right.

\begin{figure}[t]\centering
\includegraphics[width=\linewidth]{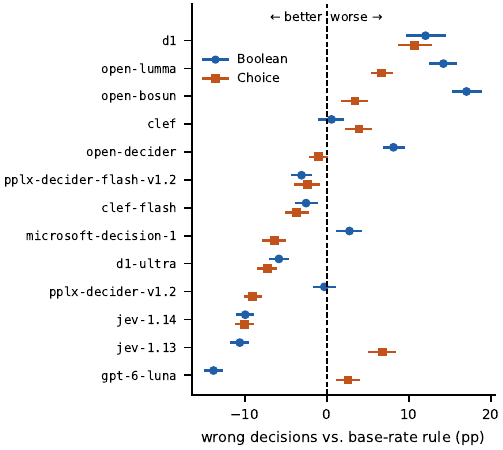}
\caption{Right of the dashed line, a type makes more wrong threshold decisions than the base-rate rule. Share of wrong
decisions minus that of the base-rate rule, in percentage points, over the matched events of the three families and
$\tau\in\{0.1,0.3,0.5,0.7,0.9\}$, with 95\% item-bootstrap intervals.}\label{fig:decisions}
\end{figure}
\begin{figure}[t]\centering
\includegraphics[width=\linewidth]{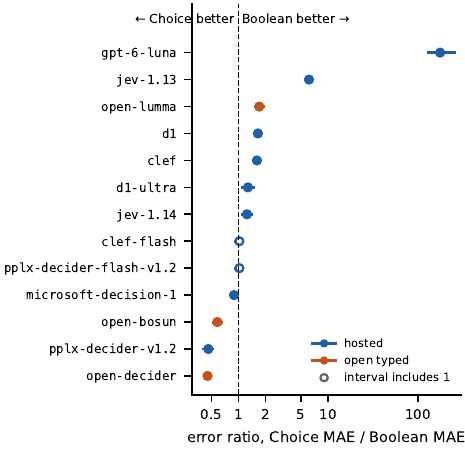}
\caption{Right of the dashed line, the Boolean is more accurate than the Choice on the four-color bags. Error ratio
(Choice MAE over Boolean MAE, log scale) with 95\% intervals from 2000 bootstrap resamples of the 99--100 items; a
hollow marker means the interval includes one.}\label{fig:panel}
\end{figure}

\paragraph{Switching type does not always help} On $\pdeNBothWorseCI$ of the $\pdeNSystems$ systems, both types make detectably more wrong decisions than the base-rate rule. They are the
top three rows of Fig.~\ref{fig:decisions}, and only \sys{d1} among them is a hosted service. Weighting each wrong
decision by its regret, how far the exact probability lies from the threshold, gives the same three systems: the
base-rate rule's mean regret is $\pdeRegretConst$ per decision, and theirs is $\pdeRegretLo$ to $\pdeRegretHi$.
Nor is either type more accurate across the panel: on the bags, some systems have the lower MAE with the Boolean and
others with the Choice (Fig.~\ref{fig:panel}).

\begin{table}[t]\centering\footnotesize\setlength{\tabcolsep}{2pt}
\begin{adjustbox}{max width=\columnwidth}\begin{tabular}{@{}lrrrr@{}}
\toprule
 & \multicolumn{2}{c}{within-event $\rho$} & \multicolumn{2}{c}{wrong \% after $T$} \\
\cmidrule(lr){2-3}\cmidrule(lr){4-5}
system & Bool. & Choice & Bool. & Choice \\
\midrule
\texttt{gpt-6-luna} & 1.00 & 0.73 & \textbf{0.1} & \textbf{9.3} \\
\texttt{jev-1.13} & 0.98 & 0.84 & \textbf{3.7} & \textbf{7.8} \\
\texttt{jev-1.14} & 0.97 & 0.97 & \textbf{3.1} & \textbf{3.4} \\
\texttt{clef} & 0.89 & 0.92 & \textbf{9.8} & \textbf{7.4} \\
\texttt{clef-flash} & 0.88 & 0.93 & \textbf{7.7} & \textbf{5.6} \\
\texttt{d1} & 0.73 & 0.88 & \textbf{9.2} & \textbf{7.4} \\
\texttt{d1-ultra} & 0.88 & 0.93 & \textbf{6.5} & \textbf{7.1} \\
\texttt{microsoft-decision-1} & 0.91 & 0.97 & 12.1 & \textbf{4.5} \\
\texttt{pplx-decider-flash-v1.2} & 0.92 & 0.93 & 12.7 & \textbf{6.2} \\
\texttt{pplx-decider-v1.2} & 0.95 & 0.97 & \textbf{9.4} & \textbf{4.6} \\
\texttt{open-decider} & 0.63 & 0.65 & 19.4 & \textbf{8.8} \\
\texttt{open-bosun} & 0.37 & 0.51 & 27.3 & \textbf{11.4} \\
\texttt{open-lumma} & \ensuremath{{-}}0.02 & \ensuremath{{-}}0.11 & 15.3 & 13.7 \\
\bottomrule
\end{tabular}
\end{adjustbox}
\caption{Information and calibration: mean rank correlation of each type with the exact probability within one event
across items (the 13 matched events of the three families), and the share of wrong threshold decisions on the test half of the bags
after a temperature fitted on the development half, in bold where it is detectably below the uniform constant's \pinConstWrong\%.}\label{tab:inform}
\end{table}
\begin{table}[t]\centering\footnotesize\setlength{\tabcolsep}{2pt}
\begin{adjustbox}{max width=\columnwidth}\begin{tabular}{@{}lrrll@{}}
\toprule
 & \multicolumn{2}{c}{wrong \%} & \multicolumn{2}{c}{minus base rate (points)} \\
\cmidrule(lr){2-3}\cmidrule(lr){4-5}
system & Bool. & Choice & Bool. & Choice \\
\midrule
\texttt{microsoft-decision-1} & 11.3 & 8.3 & \ensuremath{-}1.5 [\ensuremath{-}3.5, +0.5] & \textbf{\ensuremath{-}4.5} [\ensuremath{-}6.5, \ensuremath{-}2.5] \\
\texttt{jev-1.13} & 3.8 & 18.5 & \textbf{\ensuremath{-}9.0} [\ensuremath{-}10.8, \ensuremath{-}7.2] & \textbf{+5.7} [+3.2, +8.2] \\
\texttt{jev-1.14} & 2.7 & 3.7 & \textbf{\ensuremath{-}10.1} [\ensuremath{-}11.9, \ensuremath{-}8.4] & \textbf{\ensuremath{-}9.1} [\ensuremath{-}11.0, \ensuremath{-}7.3] \\
\texttt{clef} & 10.3 & 15.8 & \textbf{\ensuremath{-}2.5} [\ensuremath{-}4.4, \ensuremath{-}0.7] & \textbf{+3.0} [+0.5, +5.5] \\
\texttt{clef-flash} & 8.6 & 9.3 & \textbf{\ensuremath{-}4.2} [\ensuremath{-}6.4, \ensuremath{-}2.1] & \textbf{\ensuremath{-}3.5} [\ensuremath{-}5.7, \ensuremath{-}1.4] \\
\texttt{pplx-decider-v1.2} & 12.0 & 6.7 & \ensuremath{-}0.8 [\ensuremath{-}3.0, +1.4] & \textbf{\ensuremath{-}6.1} [\ensuremath{-}8.0, \ensuremath{-}4.3] \\
\texttt{pplx-decider-flash-v1.2} & 12.1 & 9.4 & \ensuremath{-}0.7 [\ensuremath{-}2.9, +1.5] & \textbf{\ensuremath{-}3.4} [\ensuremath{-}5.3, \ensuremath{-}1.3] \\
\texttt{d1} & 11.9 & 18.8 & \ensuremath{-}0.9 [\ensuremath{-}3.1, +1.3] & \textbf{+6.0} [+3.7, +8.6] \\
\texttt{d1-ultra} & 3.9 & 8.2 & \textbf{\ensuremath{-}8.9} [\ensuremath{-}10.7, \ensuremath{-}7.1] & \textbf{\ensuremath{-}4.6} [\ensuremath{-}6.7, \ensuremath{-}2.5] \\
\bottomrule
\end{tabular}
\end{adjustbox}
\caption{Bags with unequal base rates, test half ($\paNTest$ bags): share of wrong threshold decisions of each type, and
its difference from a base rate estimated on the development half ($\paRWrong$\%), with 95\% bag-bootstrap intervals,
in bold where the interval excludes zero.}\label{tab:asym}
\end{table}

\paragraph{Following the input is not enough} A type can rank items in the right order and still put its probabilities on the wrong scale, for example too close to
zero or one. Thresholds then cut it in the wrong places. Most types that lose to the base rate still follow the exact
probability. Only the answers of \pinNoInfoNames{} are nearly uncorrelated with it. Both types of every hosted system follow it, with a rank correlation of at least $\pinHostedRhoMin$
(Table~\ref{tab:inform}). After a temperature fitted on half of the bags, every system except \pinNotBeatNames{} has a type that makes fewer wrong
decisions than the base-rate rule on the other half.

\paragraph{The baseline matters} The main bags have a uniform base rate by construction. On the second set of bags,
with unequal colors, the base rate is estimated on half of the bags and decisions are scored on the other half. No
OpenRouter system loses to it with both types, but the Choice of $\paNChoiceWorse$ systems still does
(Table~\ref{tab:asym}).

\subsection{A2: Can a measurement be reused?}

A caller that has measured a type once would like to reuse the result on its next task, after the next model update
and on the next run. This section asks whether the more accurate type stays the same.

\paragraph{Across runs} \sys{d1-ultra} shows what can change under one model identifier. Its first run favors the
Boolean on the bags. Asked the same items twice more, about $\pdDays$ days later, it shows no detectable difference in one run and
favors the Choice in the other~\citep[cf.][]{atil2025nondeterminism}; from the first run to that one, the gap between
its Choice and Boolean MAE moves by \pdUltraGapChangeRep{}. Even its two later runs differ in $\pdDecUltraRepeat$\% of threshold decisions. A caller who chose the Boolean for \sys{d1-ultra} after
its first run would have kept a type that neither later run favors. Most of the other OpenRouter systems keep every ordering
(Table~\ref{tab:remeasure}).

\begin{table}[t]\centering\footnotesize\setlength{\tabcolsep}{2pt}
\begin{tabular}{@{}lrrccc@{}}
\toprule
 & mean & decisions & \multicolumn{3}{c}{more accurate type} \\
\cmidrule(lr){4-6}
system & $|\Delta|$ & changed \% & bags & histograms & chat logs \\
\midrule
\texttt{microsoft-decision-1} & 0.010 & 1.2 & C$\to$C & C$\to$C & C$\to$C \\
\texttt{jev-1.13} & 0.009 & 0.9 & B$\to$B & B$\to$B & B$\to$B \\
\texttt{jev-1.14} & 0.010 & 1.1 & B$\to$B & =$\to$= & =$\to$= \\
\texttt{clef} & 0.006 & 0.9 & B$\to$B & B$\to$B & =$\to$= \\
\texttt{clef-flash} & 0.002 & 0.2 & =$\to$= & B$\to$B & C$\to$C \\
\texttt{pplx-decider-v1.2} & 0.022 & 2.6 & C$\to$C$\to$C & C$\to$C$\to$C & C$\to$C$\to$C \\
\texttt{pplx-decider-flash-v1.2} & 0.019 & 2.4 & =$\to$=$\to$= & B$\to$B$\to$B & =$\to$C$\to$C \\
\texttt{d1} & 0.027 & 3.1 & B$\to$B$\to$B & B$\to$B$\to$B & C$\to$C$\to$C \\
\texttt{d1-ultra} & 0.065 & 6.5 & B$\to$=$\to$C & C$\to$C$\to$C & C$\to$C$\to$C \\
\bottomrule
\end{tabular}

\caption{Re-measurement under unchanged model identifiers: mean absolute change of an answer, and share of threshold
decisions of both types that change, between the first and the second run; the more accurate type per family in each
run (as in Table~\ref{tab:families}).}\label{tab:remeasure}
\end{table}

\begin{table*}[t]\centering\footnotesize\setlength{\tabcolsep}{2pt}
\begin{tabular}{@{}lrrcrrcrrc@{}}
\toprule
 & \multicolumn{3}{c}{bags} & \multicolumn{3}{c}{histograms} & \multicolumn{3}{c}{chat logs} \\
\cmidrule(lr){2-4}\cmidrule(lr){5-7}\cmidrule(lr){8-10}
system & Bool. & Choice & & Bool. & Choice & & Bool. & Choice & \\
\midrule
\texttt{gpt-6-luna} & 0.001 & 0.161 & B & 0.000 & 0.160 & B & 0.000 & 0.154 & B \\
\texttt{jev-1.13} & 0.034 & 0.210 & B & 0.042 & 0.203 & B & 0.033 & 0.195 & B \\
\texttt{jev-1.14} & 0.030 & 0.037 & B & 0.038 & 0.040 & = & 0.041 & 0.042 & = \\
\texttt{clef} & 0.099 & 0.158 & B & 0.149 & 0.183 & B & 0.163 & 0.165 & = \\
\texttt{clef-flash} & 0.092 & 0.093 & = & 0.085 & 0.154 & B & 0.143 & 0.057 & C \\
\texttt{d1} & 0.136 & 0.224 & B & 0.178 & 0.235 & B & 0.399 & 0.262 & C \\
\texttt{d1-ultra} & 0.056 & 0.071 & B & 0.091 & 0.062 & C & 0.094 & 0.065 & C \\
\texttt{microsoft-decision-1} & 0.122 & 0.109 & C & 0.155 & 0.079 & C & 0.191 & 0.040 & C \\
\texttt{pplx-decider-flash-v1.2} & 0.119 & 0.122 & = & 0.083 & 0.111 & B & 0.114 & 0.108 & = \\
\texttt{pplx-decider-v1.2} & 0.117 & 0.054 & C & 0.095 & 0.050 & C & 0.162 & 0.036 & C \\
\texttt{open-decider} & 0.200 & 0.091 & C & 0.289 & 0.134 & C & 0.183 & 0.149 & C \\
\texttt{open-bosun} & 0.252 & 0.146 & C & 0.407 & 0.133 & C & 0.246 & 0.206 & C \\
\texttt{open-lumma} & 0.161 & 0.273 & B & 0.376 & 0.186 & C & 0.325 & 0.158 & C \\
\bottomrule
\end{tabular}

\caption{MAE of the asked Booleans and of the same events derived from the Choice, per task family, and the more
accurate type where the 95\% interval of the ratio of the two (2000 bootstrap resamples of items) excludes one (B:
Boolean, C: Choice, =: not resolved). On histograms and chat logs the events are cumulative.}\label{tab:families}
\end{table*}

\paragraph{Across tasks and versions} The more accurate type also changes with the task. On $\pmfNChange$ of the $\pmfNSystems$ systems, the Boolean is more accurate on one task family and the Choice on
another (Table~\ref{tab:families}). \sys{d1}, for example, has the more accurate Boolean on the bags and the more
accurate Choice on the chat logs. A caller who measured it on one family would pick the wrong type for the other. The bags ask about single colors and the other families about cumulative events, so the task family
and the event form change together. The gap between the types changes with the version. \sys{jev-1.13} and \sys{jev-1.14} answered the same items and have Choice errors of $\npanelJevThirteenChoice$ and
$\npanelJevFourteenChoice$ on the bags, while their Booleans stay similar.

\subsection{A3: Do two label-free tools replace labels?}

Labels from the caller's own task cost effort, so a check or repair that needs none is attractive. This section asks whether the partition check picks the better type on the bags and on ChaosNLI. It then asks whether the
rebuilt Choice, which helps on the bags, helps on human judgments too.

\paragraph{The partition check warns but does not beat always using the Boolean} Exactly one color is drawn, so the
four per-color answers should sum to one. When \sys{gpt-6-luna} is asked each color as its own two-way Choice, the answers miss a sum of one by $\ppbLunaDevC$ on
average, against $\ppbLunaDevB$ for its Booleans, and the two-way Choices are indeed the inaccurate type. As a selector,
the check does not detectably improve on always using the Boolean: on the $\pnNSystems$ OpenRouter systems, the selection has a
pooled MAE of $\pbnSel$ against $\pbnB$ on the bags (95\% interval of the difference $\pbnSelMinusB$) and $\pnSelMae$
against $\pnBoolMae$ on ChaosNLI ($\pnSelMinusBool$). An oracle that gives each system its more accurate form bounds
what a selector could gain. On the bags it reaches $\pbnOrc$ ($\pbnOrcMinusB$ against the Boolean), because the
two-way Choice is the more accurate form for only $\pbnNBetterChoice$ of the $\pnNSystems$ systems. On ChaosNLI it
reaches $\pnOrcMae$ ($\pnOrcMinusBool$), a gain the check does not find. The check still works
as an alarm. By the bound, answers that miss a sum of one by much cannot be accurate, but answers that sum to one can
still be inaccurate.

\begin{figure}[t]\centering
\includegraphics[width=\linewidth]{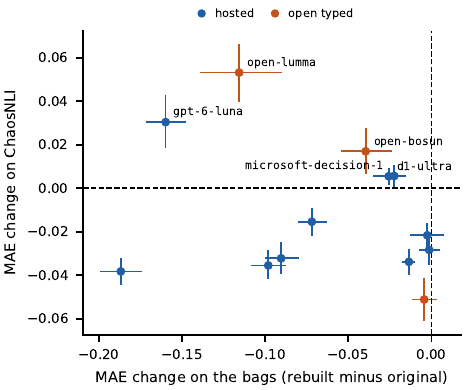}
\caption{Left of the vertical line, the rebuilt Choice is more accurate than the original Choice on the bags; below the
horizontal line, it is more accurate on ChaosNLI. MAE change (rebuilt minus original) against the exact probability on
the bags and against the annotator distribution on the ChaosNLI test half, with 95\% paired intervals. Labels mark the systems for which it is detectably less accurate on ChaosNLI.}\label{fig:repair}
\end{figure}

\paragraph{A repair that works on the bags can hurt on human labels} On the bags, the rebuilt Choice lowers
the MAE of $\prsNBagHelpsCI$ of $\prsNSystems$ systems and raises it for none. On ChaosNLI it raises the MAE for
$\prsNHurts$ systems, among them \sys{gpt-6-luna} (Fig.~\ref{fig:repair}). All $\prsNHurts$ systems it hurts on human labels are among those it helps on the bags. For $\prsNHurtBoolWorse$ of these $\prsNHurts$ systems the loss comes with the change of type: on the ChaosNLI test half
their per-label Booleans as returned are detectably less accurate than the original Choice, and dividing by the sum
does not detectably raise their MAE. For the fifth, \sys{microsoft-decision-1}, the two types are not resolved and the
division detectably raises the MAE. \sys{gpt-6-luna}, whose Boolean is the more accurate type on all three synthetic
families, has a Boolean MAE of $\prsLunaBool$ against $\prsLunaRaw$ for its Choice on that half. Labels from the target task help instead: a temperature fitted on the development half lowers the
pooled MAE of the original Choice from $\prsRaw$ to $\prsTemp$, and lowers it detectably for $\prsNTempLowersCI$ of
the $\prsNSystems$ systems~\citep[cf.][]{wang2022capture}.

\section{Related work}

Calibration of language models depends on the question format~\citep{kadavath2022know}, and option
order~\citep{arxiv2309_03882} and multiple-choice-to-True/False conversion~\citep{zhou2024selfconsistency} change
generative models' answers; the elicitation channel changes the reported probability~\citep{tian2023justask}.
\citet{arxiv2610_19536} report that, for every typed model they test on tabular data, at least a third of the row decisions
change under some rewrite of the table or the question. Calibration against a majority label is problematic when annotators
disagree~\citep{baan2022stop}, which is why we score against label distributions. Hosted models can return different
outputs under settings meant to be deterministic~\citep{atil2025nondeterminism}. Closer to our setting,
\citet{arxiv2610_17016} build simulated text domains with exact class probabilities and score 16 models, several of
them decision models. \citet{arxiv2610_24692} compare a typed model with input-independent baselines in clinical
triage, where its negative log-likelihood lies below that of a uniform prediction but above that of the development
prior. Rebuilding a Choice from per-option Booleans has been tested against gold labels: it raises the retained accuracy
of a hosted typed model in medicine, though not of every model~\citep{arxiv2610_18529}. Scoring each option with
its own yes/no question picks the right intent or topic less often~\citep{arxiv2610_16655}. \citet{arxiv2610_38623}
find that discrepancies between a model's question types, including between a Choice and normalized per-option
Booleans, predict decision failure, and reduce them with a trained calibrator. Our partition check is a label-free
warning of this kind, and we also test it as a selector. On typed decision interfaces, one
confidence threshold handles a smaller share of requests in every other language than in English~\citep{yao2026threshold}.
A confidence monitor built on a typed model's output probabilities misses most hijacks from a plain injected note, at a
threshold that alarms on at most 5\% of clean answers~\citep{yao2026hijack}. We compare the decisions of both types
with the base rate on $\npanelNModels$ systems, including two versions of one model. We re-measure nine hosted systems
under unchanged identifiers, and test whether a label-free check and a label-free repair carry from exact targets to
human label distributions.

\section{Conclusion}

A measurement made in one setting does not always carry over to the next. A caller who reads typed decision
probabilities should therefore score the type it uses on its own task, against a reference and
a constant, and score it again when the version or the run changes. Each part of this advice answers one of the
failures above. Scoring the type is needed because neither type is reliably more accurate, and both can lose to the
base rate even when their answers follow the input (A1). Scoring on the caller's own task, and again after a change, is needed because the more accurate type changes with the
task family and, for one hosted system, between runs under the same identifier. The gap between the types also changes
with the version (A2). The two label-free tools we tested do not replace those labels. In choosing between per-option forms, the partition check does not detectably improve on always using the Boolean. A repair that helps on the synthetic bags can hurt on human labels, where a temperature
fitted on labeled data helps (A3). Without labels, the check can still warn: a large deviation from summing to one means the
type has a large error.

\section*{Limitations}

The states are short English texts. The exact-probability tasks are stated-count arithmetic with 100 items per family.
The human-label task is one natural language inference dataset, whose annotator distributions measure disagreement
rather than a posterior. The hosted systems are the identifiers listed in Appendix~\ref{app:protocol}; their architectures
and training recipes are undocumented, so the measurements describe their behavior as served. The re-measurement cannot separate a change of the model from a change of the serving stack.

\section*{GenAI Usage Disclosure}
Claude Code (Anthropic) was used to implement the experiments and analyses and to draft and revise the manuscript.
The authors directed each step of the study, read and revised the manuscript, verified every reported number against
the underlying data, and are responsible for all of its content.

\section*{Ethical Considerations}
The study queries publicly released models and services as they are offered and uses synthetic tasks and the public
ChaosNLI dataset. It involves no human participants and no personal data.


\appendix
\section{Protocol}\label{app:protocol}
\sloppy

{\raggedright \paragraph{Systems and runs}\label{app:systems} \sys{gpt-6-luna} is called through OpenAI's Decisions API,
and the other nine hosted systems through OpenRouter as \mbox{\texttt{typesafe/jev-1.13}}, \mbox{\texttt{typesafe/jev-1.14}}, \mbox{\texttt{cloudflare/clef}}, \mbox{\texttt{cloudflare/clef-flash}}, \mbox{\texttt{perplexity/pplx-decider-v1.2}}, \mbox{\texttt{perplexity/pplx-decider-flash-v1.2}}, \mbox{\texttt{liquid/d1}}, \mbox{\texttt{liquid/d1-ultra}} and
\mbox{\texttt{microsoft/microsoft-decision-1}}. OpenRouter sent \sys{clef} to DeepInfra, \sys{clef-flash} to Makora and every
other system to its maker, except that \sys{pplx-decider-v1.2} went to Makora in every run but the first run and the main
ChaosNLI run, which Perplexity served. Each main request carries all questions about a state, except for \sys{open-bosun}, whose
code takes one question per prompt. \sys{gpt-6-luna} refused the four-way Choice of one bag, which is left out of its bag
measures. The re-measurement resends the main requests of the three families unchanged.\par}

\paragraph{Synthetic items} A seeded generator draws the 300 evaluation items (297 distinct states). Weights come from
a symmetric Dirichlet distribution whose concentration is drawn uniformly from $\{0.5, 1, 3\}$, and the counts follow
the weights times the total, each at least 1. A bag draws its total uniformly from $\{20, 40, 50, 100\}$ and states the
sum of its four counts as its total; a histogram has 100 reviews over five star levels and a chat log 100 chats over
first-anger turns 1--5 and never. A bag's request has nine questions: the four per-color Booleans, the four-way Choice
(``What color is the drawn token?''), a negated Boolean, a two-way Choice between one color and the rest, a Choice
over two color groups and a union Boolean. A histogram asks ``Does the drawn review have at least $k$ stars?''
($k=2,\ldots,5$) and a Choice over the five ratings, among other questions; a chat log asks ``In the drawn chat, had the customer expressed
anger by the end of turn $k$?'' ($k=1,\ldots,5$) and a Choice over the first-anger turn (turns 1--5 or never).

\paragraph{Unequal base rates} Another seeded generator draws $\paNBags$ bags with totals as above. Their four weights are gamma draws with shapes
$\paAlpha$ for red, green, blue and yellow, the counts sum exactly to the total, and no color exceeds 85\% of a bag; no state repeats another bag. They use the main
bag request. The even-numbered bags estimate the base rate and the odd-numbered bags ($\paNTest$) are scored.

\paragraph{Bag partition check} Two sets of 80 bags give 320 per-color events each: 80 of the main evaluation bags,
and new bags whose counts sum exactly to their total, with no color above 85\%, and whose states appear in no other bag set. Each event gets three one-question requests:
a Boolean and a two-way Choice in each option order, with the two orders averaged. All three use the question ``Consider the proposition: the drawn token
is $X$.'' and the criteria ``The drawn token is $X$.'' and ``The drawn token is a colour other than $X$.'', which are
also the Choice's options. Intervals resample bags. The selector comparison uses both sets, 160 bags per system. The
deviations reported for \sys{gpt-6-luna} are from the new bags; on the main bags they are $\ppbLunaDevCSecond$ and
$\ppbLunaDevBSecond$.

\paragraph{ChaosNLI} The items are the first $\pfNItems$ of a fixed random permutation of the Hugging Face dataset
\path{tasksource/chaos-mnli-ambiguity}~\citep{hf2023chaosmnli}, a repackaging of ChaosNLI, which is distributed
under the CC BY-NC 4.0 license and builds on MNLI~\citep{williams2018mnli}. The state is the premise and the hypothesis. The
Choice asks ``How does the hypothesis relate to the premise?'' with the options entailment (``the hypothesis definitely
follows from the premise''), neutral (``the hypothesis might or might not be true given the premise'') and
contradiction (``the hypothesis definitely contradicts the premise''); each label's Boolean asks ``Does the hypothesis
relate to the premise in this way: $\langle$description$\rangle$?''. An item is in the development half ($\prsNDev$
items) when the CRC-32 checksum of its id is even, and in the test half ($\prsNTest$) otherwise. For the partition
check, each system answers every item twice in one session: once with the main request and once with three per-label
two-way Choices, each with that label's Boolean question and its two criteria texts as options. Deviation and MAE
average the per-label answers over items. Paired intervals use $\pfNBoot$ bootstrap resamples of items; pooled
intervals resample all $\pnNItems$ items jointly across systems.

\paragraph{Temperature} A temperature $T$ rescales a type's answers: $T>1$ moves them toward the middle and $T<1$
toward the extremes; at $T=2$, a Boolean answer of $0.9$ becomes $0.75$. Formally, the Boolean's log-odds and the Choice's
log-probabilities are divided by $T$, after answers below $10^{-4}$ (and, for a Boolean, above $1-10^{-4}$) are set to that bound. $T$ is fitted to the MAE on
one half of the items by a bounded search over $\log T\in[-2,3]$ and scored on the other half: the
even- and odd-numbered main bags, and the development and test items of ChaosNLI.

\end{document}